\documentclass[
  a4paper,
  fontsize=12pt,
  captions=tableheading,
  parskip=never,
  ]{scrartcl}
\setkomafont{caption}{\small}
\setkomafont{captionlabel}{\bfseries}
\addtokomafont{disposition}{\rmfamily\boldmath}
\addtokomafont{publishers}{\large}
\addtokomafont{subject}{\mdseries\large}

\pdfoutput=1

\usepackage{graphicx}

\usepackage[utf8]{inputenx}
\usepackage[T1]{fontenc}
\usepackage[british]{babel}
\usepackage{csquotes}

\usepackage{amsmath}
\usepackage{amsfonts}
\usepackage{amssymb}

\usepackage{subcaption}
\usepackage{import}
\usepackage{xspace}
\usepackage{graphicx}
\usepackage[shortcuts]{extdash}
\usepackage{siunitx}
\DeclareSIUnit \lightspeed {\text{{c}}}
\usepackage{xcolor}
\definecolor{linkblue}{HTML}{264772}
\usepackage{hyperref}
\hypersetup{
    colorlinks=true,
    linktocpage=true,
    linkcolor=linkblue,
    citecolor=linkblue,
    urlcolor=linkblue
}
\usepackage[noabbrev,capitalise]{cleveref}

\usepackage[backend=biber, style=numeric-comp, sorting=none]{biblatex}
\DeclareDataInheritance{mvbook}{book}{\noinherit{volumes}}
\AtEveryBibitem{%
  \iffieldundef{doi}{}{\clearfield{url}} 
  \clearfield{urlyear} %
}

\graphicspath{ {./pictures/} }
\DeclareGraphicsExtensions{.pdf,.png,.jpg}

\newcommand{\dEdx}{\ensuremath{\mathrm{d}E/\mathrm{d}x}\xspace}

\newcommand{\orderof}[1]{\ensuremath{\mathcal{O}(#1)}\xspace}
\newcommand{\rphi}{\ensuremath{r\varphi}\xspace}

\begin{document}

\hyphenation{
  am-pli-fi-ca-tion
  col-lab-o-ra-tion
  per-for-mance
  sat-u-rat-ed
  se-lect-ed
  spec-i-fied
}

\title{Recent Performance Studies of the GEM-based TPC Readout \protect \\(DESY Module)}
\author{Ties Behnke, Ralf Diener, Ulrich Einhaus, Uwe Kr\"amer,
  \\ Paul Malek, Oliver Sch\"afer, Mengqing Wu}
\subject{Talk presented at the International Workshop on Future Linear Colliders (LCWS2019), Sendai, Japan, October 30th 2019.}
\publishers{Deutsches Elektronen\-/Synchrotron, DESY}
\date{}

\maketitle

\begin{abstract}
  \noindent
  For the International Large Detector (ILD) at the planned International Linear Collider (ILC) a Time Projection Chamber (TPC) is foreseen as the main tracking detector.
  To achieve the required point resolution, Micro\-/Pattern Gaseous Detectors (MPGD) will be used in the amplification stage.
  A readout module using a stack of three Gas Electron Multipliers (GEM) for gas amplification was developed at DESY and tested at the DESY II Test Beam Facility. 

  After introducing the readout module and the infrastructure at the test beam facility, the performance related to single point and double\-/hit resolution of three of these modules is presented. This is followed by results on the particle identification capabilities of the system, using the specific energy loss {\dEdx}, and simulation studies, aimed to investigate and quantify the impact of high granularity on {\dEdx} resolution.

 In addition, a new and improved TPC field cage and the LYCORIS Large\-/Area Silicon\-/Strip Telescope for the test beam are described. The LYCORIS beam telescope is foreseen to provide a precise reference of the particle trajectory to validate the momentum resolution measured with a large TPC prototype. For this purpose, it is being installed and tested at the test beam facility within the so\-/called PCMAG (Persistent Current Magnet). 
 
\end{abstract}

\clearpage

\tableofcontents
\newpage
\section{Introduction}
This project is part of the design effort for the International Large Detector (ILD) at the future International Linear Collider experiment (ILC)~\cite{ilc_tdr_detectors}. The FLC-TPC group at DESY is part of the LCTPC collaboration~\cite{LCTPC}, which is driving the efforts of developing a TPC for the ILD. Different readout and amplification technologies based on Micro\-/Pattern Gaseous Detectors (MPGD) are currently studied within the collaboration. FLC-TPC is committed to designing a self-supporting readout composed of GEMs~\cite{GEM} mounted on thin ceramic grids~\cite{1748-0221-8-12-P12009}.

\section{The LCTPC Test Beam at DESY}
\subsection{The DESY GridGEM Module}
\begin{figure}[!bp]
  \begin{subfigure}{0.5\textwidth}
    \centering
    \includegraphics[width=\textwidth,height=0.33\textheight,keepaspectratio=true]{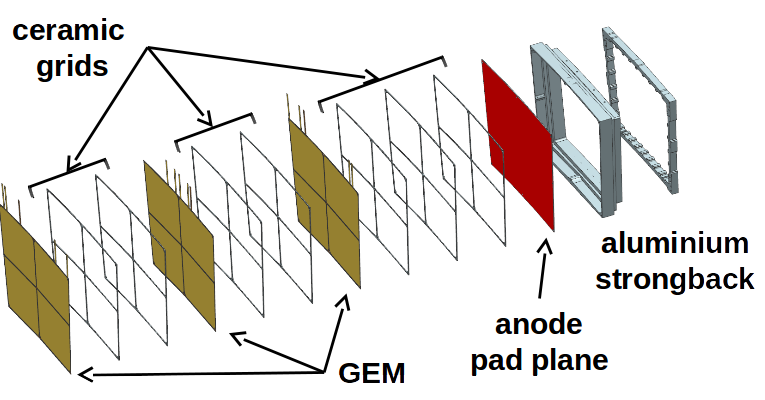}
    \caption{Explosion view of the GEM module.
      \label{fig:view_readoutmodule}}
  \end{subfigure}%
  \begin{subfigure}{0.5\textwidth}
    \centering
    \includegraphics[width=\textwidth,height=0.185\textheight,keepaspectratio=true]{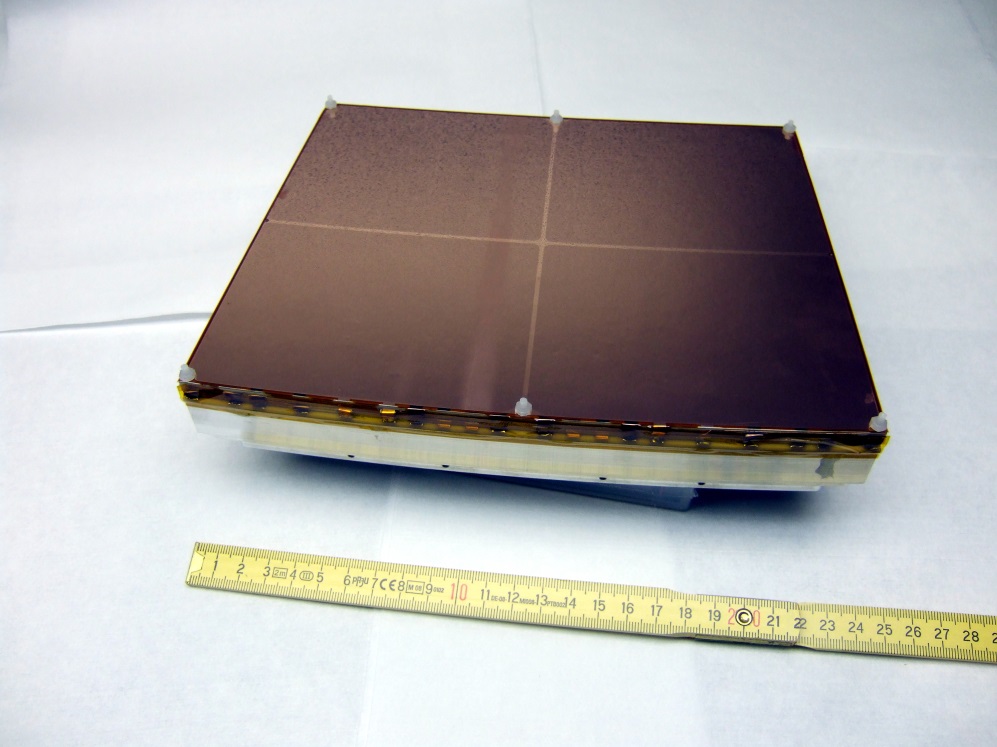}
    \caption{Picture of an assembled GEM module.
      \label{fig:GEM}}
  \end{subfigure}
  \caption{The DESY GridGEM module.}
\end{figure}

In order to maximize the sensitive area while minimizing the gaps and the material budget, the DESY GridGEM module has been developed.
The mechanical base of the module is an aluminium back frame that is used to mount the module in the TPC prototype end plate.
Onto this a printed circuit board (PCB) is glued, which contains the readout pad plane.
This structure also provides the gas tightness of the module.
The pad plane contains 4832 readout pads with a pitch of \SI{1.26}{\mm} arranged in 28 concentric rows with a pitch of \SI{5.85}{\mm} in radial direction.
On top of the pad board the amplification structure is built.
It consists of three GEMs supported by \SI{1}{\mm} high alumina\-/ceramic frames.
Additional frames are used as spacers between the GEM foils to define transfer and induction gaps, as shown in \cref{fig:view_readoutmodule}.
A fully assembled module is shown in \cref{fig:GEM}.
Since the frame bars are only \SI{\sim1.4}{\mm} wide, the modules have an active area of about \SI{95}{\percent}.
The data is read out by a modified version of the ALICE TPC Readout (ALTRO)~\cite{ALTROchip,EudetPCA16}.

\subsection{The LCTPC Setup at the DESY II Test Beam Facility}
The LCTPC readout prototypes are tested in an electron beam at the DESY II Test Beam Facility~\cite{Diener:2018qap}.
The beam line delivers electrons in the momentum range from \SIrange{1}{6}{\GeV\per\lightspeed} at rates from \orderof{\si{\kilo\Hz}} to \SI{\sim100}{\Hz}, respectively.

\begin{figure}[!bp]
  \begin{subfigure}{0.5\textwidth}
    \centering
    \includegraphics[width=\textwidth,height=0.33\textheight,keepaspectratio=true]{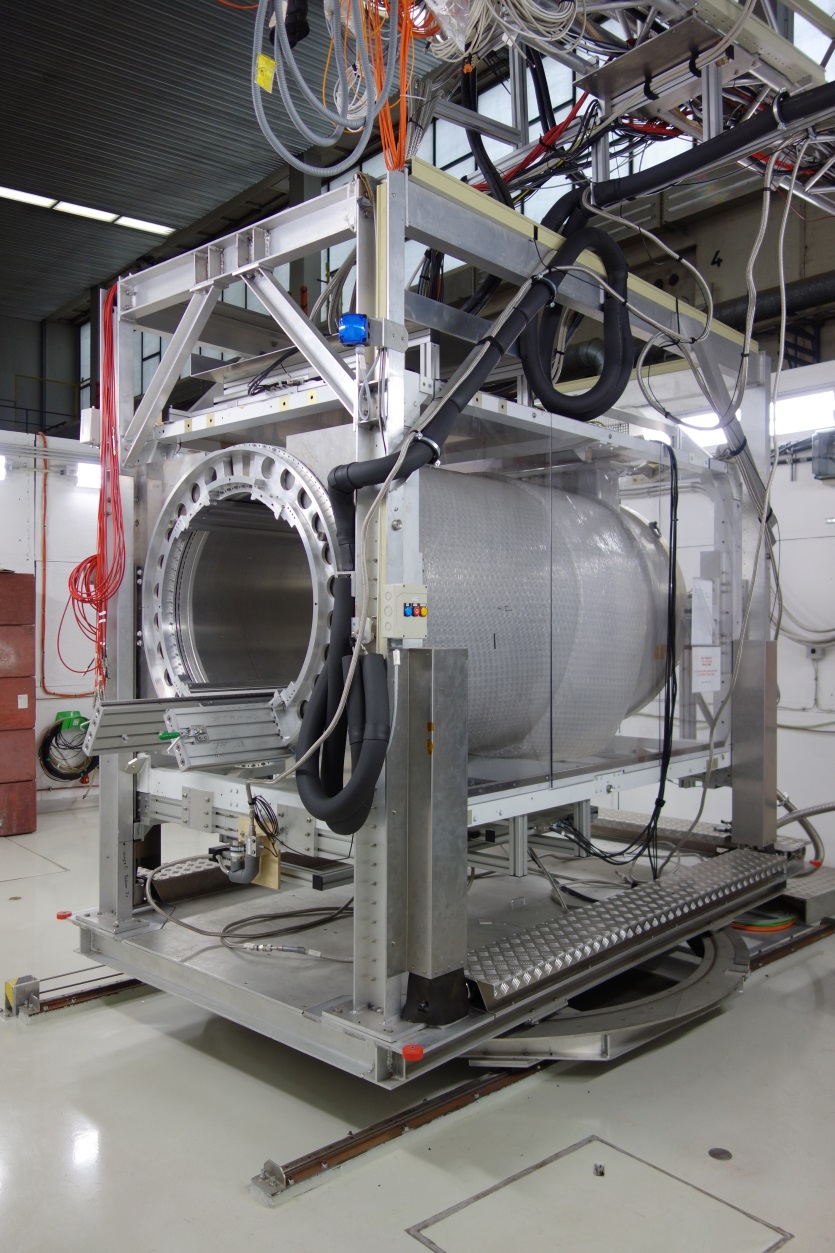}
    \caption{The PCMAG.
      \label{fig:PCMAG}}
  \end{subfigure}%
  \begin{subfigure}{0.5\textwidth}
    \centering
    \includegraphics[width=\textwidth,height=0.33\textheight,keepaspectratio=true]{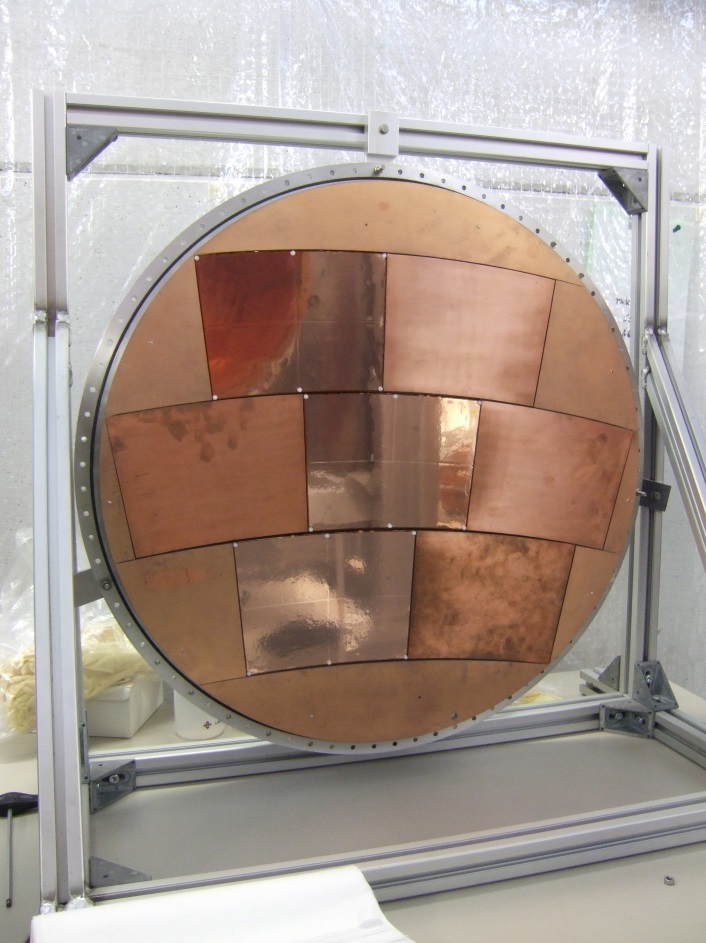}
    \caption{The LPTPC anode end-plate~\cite{FMueller2017}.
      \label{fig:LPTPC}}
  \end{subfigure}%
  \caption{The PCMAG solenoid and the LPTPC setup at the DESY II Test Beam Facility.}
\end{figure}

The setup consists of a large TPC prototype (LPTPC) and a superconducting solenoid magnet, the so\-/called Persistent Current Magnet (PCMAG), shown in \cref{fig:PCMAG}. A coincidence signal from four scintillator planes placed in the beam path in front of the PCMAG is used as the trigger signal for the main setup.
Slow control is implemented using the Distributed Object\-/Oriented Control System (DOOCS).
The solenoid produces a magnetic field of up to \SI{1}{\tesla}.
It is mounted on a movable stage installed in the experimental area T24/1 of the test beam.
The stage can be translated vertically and horizontally perpendicular to the beam axis and rotated around the vertical axis.
Additionally, the TPC can be rotated around its axis inside the magnet.
This allows to scan the beam over the full volume of the TPC and to change both the polar and azimuthal angle of the beam relative to the TPC.
The usable inner diameter of the PCMAG is \SI{85}{\cm}.
Due to a lightweight construction without a field return yoke, the magnet wall contributes only about \SI{20}{\percent} of a radiation length $X_0$~\cite{pcmag:magnet}.

The LPTPC has a maximum drift length of about \SI{570}{\mm} and an inner diameter of \SI{75}{\cm}.
The anode end plate contains slots for 7 modules, arranged in three staggered rows.
Since all module slots are of identical shape, the rows are not concentric.
Unused slots can be filled with dummy modules for a uniform termination of the drift field.
This is shown in \cref{fig:LPTPC}, where the active modules can be distinguished by their more reflective surface.

\subsection{Established System Performance}
\begin{figure}[!bp]
  \begin{subfigure}[t]{0.5\textwidth}
    \centering
    \includegraphics[width=\textwidth,height=0.25\textheight,keepaspectratio]{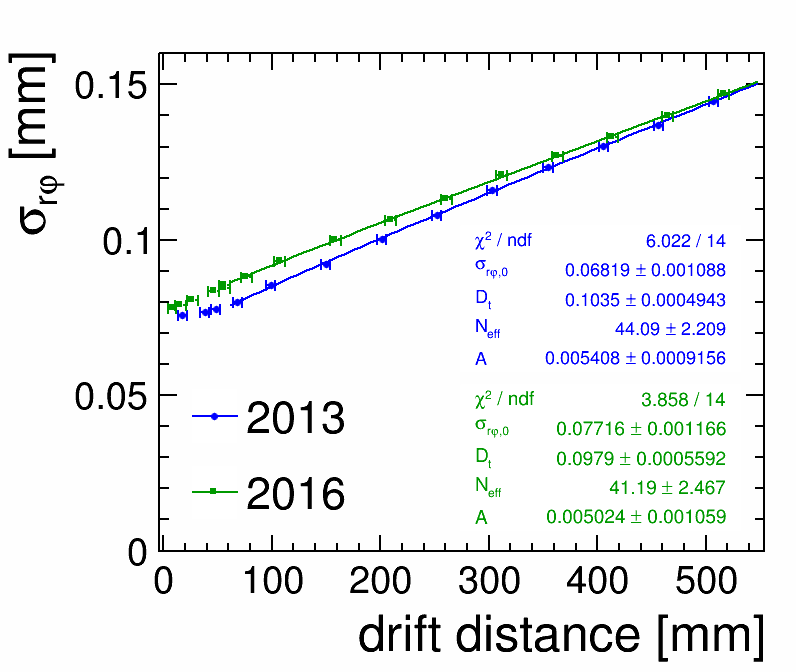}
    \caption{Comparison of measurements.}
    \label{fig:rphirescomp}
  \end{subfigure}%
  \begin{subfigure}[t]{0.5\textwidth}
    \centering
    \includegraphics[width=\textwidth,height=0.25\textheight,keepaspectratio]{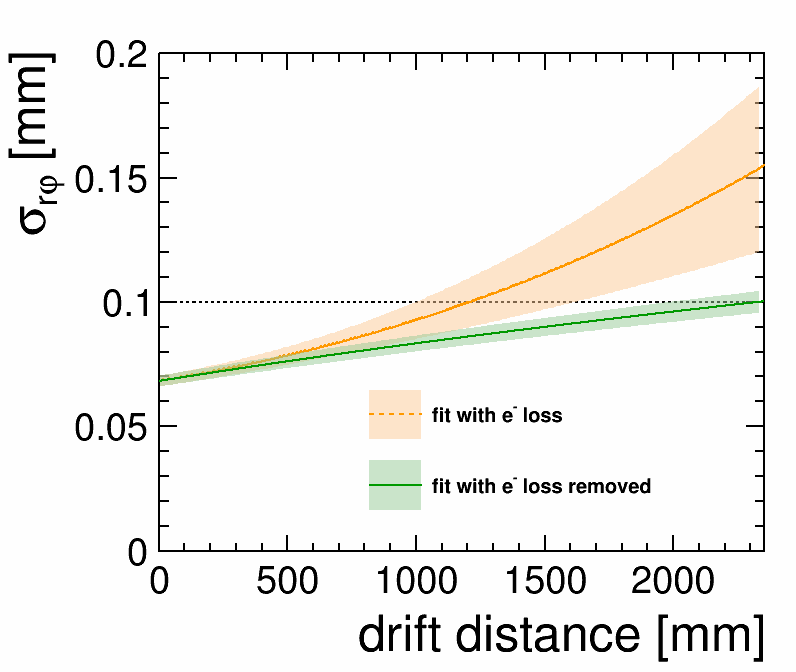}
    \caption{Extrapolation to the ILD.}
    \label{fig:rphiresextra}
  \end{subfigure}%
  \caption{The transverse resolution $\sigma_{\rphi}$
    \subref*{fig:rphirescomp}~as measured in the two beam test campaigns.
    The markers represent the data and the lines show fits of the expected behaviour as described in~\cite{Mueller:301339}.
    \subref*{fig:rphiresextra}~An extrapolation to the size and magnetic field of the ILD TPC.
    The colored bands show the $\SI{2}{\sigma}$ confidence intervals.
  }
  \label{fig:pointrescomp}
\end{figure}

In two beam test periods in 2013 and 2016, respectively, the point resolution of the system has been determined.
Generally good agreement was found between the data of both tests.
A small discrepancy was found in the \rphi\=/resolution at small drift distances, as shown in \cref{fig:rphirescomp}, visible as a larger intrinsic resolution $\sigma_{\rphi,0}$ of the system and a smaller transverse diffusion constant $D_\mathrm{t}$ in the newer data.
Various systematic effects have been studied, including gas pressure and temperature, gas composition and contamination as well as settings of high voltages and electric and magnetic fields.
Within reasonable fluctuations, none of these effects can explain the observed discrepancy and further investigations are required.

Using the measurement data, the point resolution can be extrapolated to the conditions of the ILD TPC.
Since the transverse diffusion is reduced due to the higher magnetic field of \SI{3.5}{\tesla}, the expected diffusion constant is simulated using Magboltz~\cite{Magboltz}.
\Cref{fig:rphiresextra} shows the resulting \rphi\=/resolution for the 2013 data set.
The orange line represents the direct extrapolation of the measurement, including the measured electron attachment rate.
Since the attachment rate is expected to be negligible in the ILD TPC, due to much lower gas contamination, for the green line the attachment rate was set to zero.
In this case both data sets result in a resolution around $\sigma_{\rphi}=\SI{100}{\um}$ at the full drift length of the ILD TPC, fulfilling the requirement set in~\cite{ilc_tdr_detectors}.

\begin{figure}[tbp]
  \centering
  \includegraphics[width=0.5\textwidth,height=0.33\textheight,keepaspectratio]{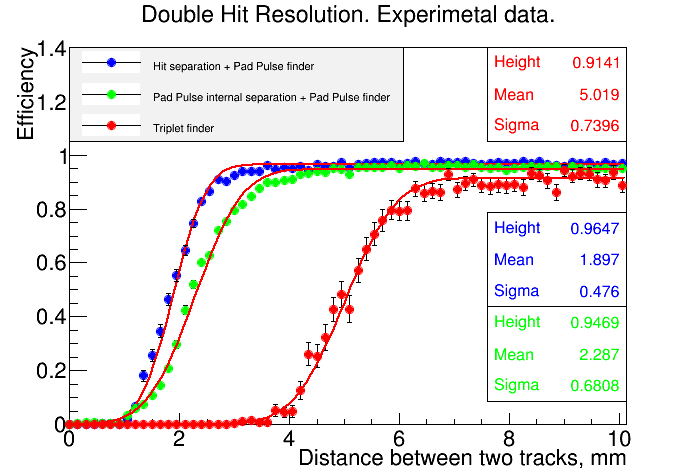}
  \caption{The two\-/hit separation efficiency versus the distance between the respective tracks.
    The three different marker color represent different steps of the algorithm.
    Red: no attempted hit separation; green: intermediate step; blue: full algorithm.
  }
  \label{fig:dhr}
\end{figure}

A related quantity is the double\-/hit resolution, which is important in the case of close\-/by or partially overlapping tracks.
Here, an algorithm was developed that first identifies potential double\-/hit candidates based on preliminary track fits.
It then performs fits to the charge distribution of both a single\-/hit and a double\-/hit hypothesis.
If the double\-/hit hypothesis performs significantly better than the single\-/hit hypothesis, the charge distribution is split into two hits based on the fit.
The blue dots in \cref{fig:dhr} show the two\-/hit separation efficiency of this algorithm versus the distance between the two reconstructed tracks at the location of the hits.
The other colors represent intermediate steps of the process.
The solid lines represent fits of an error function to the respective data.
The efficiency reaches \SI{50}{\%} at distances around and below \SI{2}{\mm}, slightly depending on the drift distance due to diffusion.
This fulfills the requirement specified in~\cite{ilc_tdr_detectors}.

\subsection{Determination of {\dEdx} Resolution}

\begin{figure}[tbp]
  \begin{subfigure}[b]{0.5\textwidth}
    \centering
    \includegraphics[width=\textwidth,height=0.25\textheight,keepaspectratio]{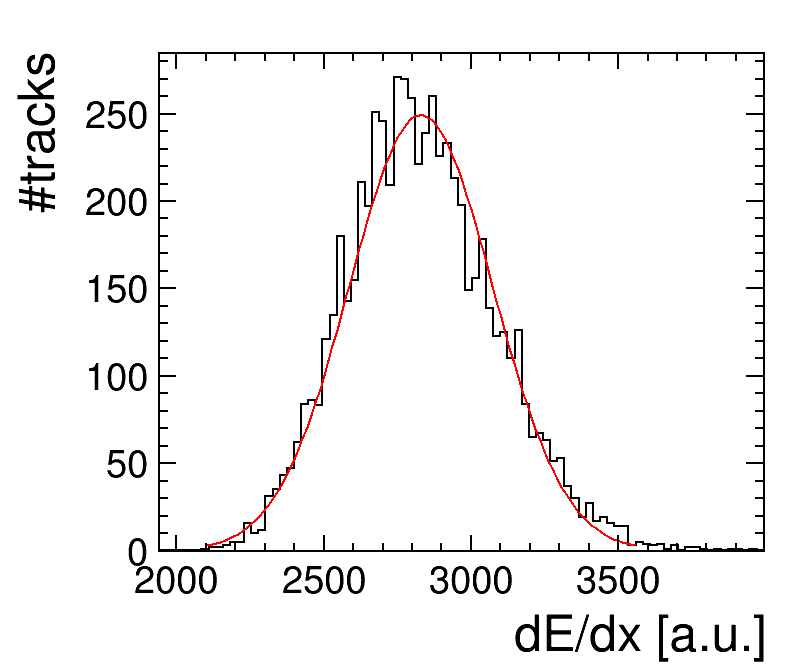}
    \caption{Track \dEdx distribution.}
    \label{fig:averagedEdx}
  \end{subfigure}%
  \begin{subfigure}[b]{0.5\textwidth}
    \centering
    \includegraphics[width=\textwidth,height=0.265\textheight,keepaspectratio]{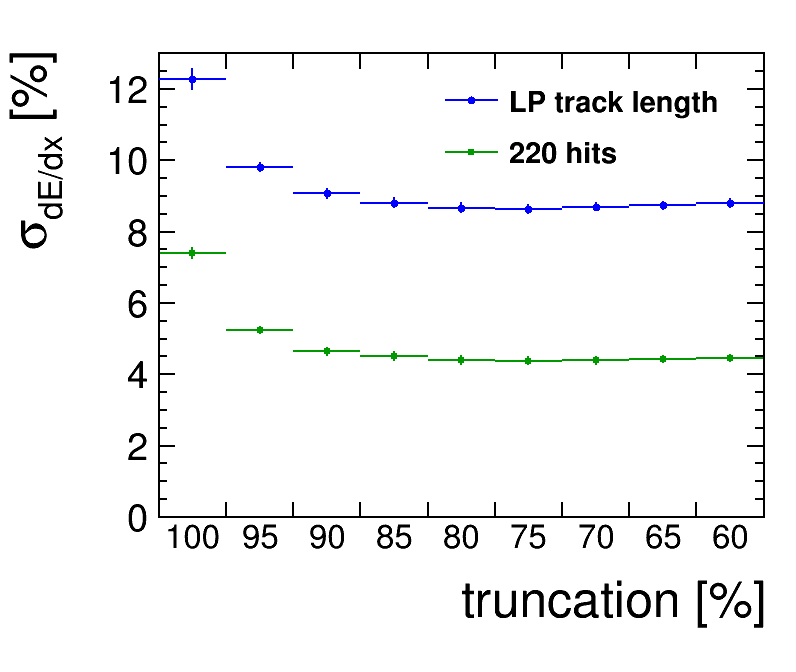}
    \vspace{-0.015\textheight}
    \caption{Optimization of the truncated mean.}
    \label{fig:truncatedmean}
  \end{subfigure}%
  \caption{The determination of the \dEdx resolution.
    \subref*{fig:averagedEdx}~The distribution of the truncated mean of \dEdx per track.
    The red line represents the Gaussian fit used to extract the resolution.
    \subref*{fig:truncatedmean}~The \dEdx resolution as a function of the truncation fraction.
    The blue marks show the resolution for the measured LPTPC tracks and the green marks an extrapolation to the full ILD TPC.
  }
  \label{fig:dEdx}
\end{figure}

The average energy loss on a track is calculated from the charge distribution of all hits that were assigned to this track by a track finding algorithm~\cite{MarlinTPC,KleinwortHough}.
On average, each track in the large TPC prototype contains about 55 hits that are considered valid for the \dEdx calculation.
The mean {\dEdx} per track in the LPTPC is shown in \cref{fig:averagedEdx}.
The relative resolution is extracted from a Gaussian fit as $\sigma$ divided by the mean.
To achieve a stable estimate of the average energy loss from the Landau\-/like primary ionization distribution the mean of a truncated distribution is calculated, where only a certain fraction of the lowest charge hits is taken into account.
As shown in \cref{fig:truncatedmean}, a shallow optimum of the resolution is found around a truncation fraction of \SI{75}{\percent}.
For tracks extrapolated to the full ILD TPC as described below the same optimum is found.
Using the optimal truncation fraction the resolution for the measured tracks is $\sigma_{\dEdx}=\SI{8.62(14)}{\percent}$.

\begin{figure}[tbp]
    \centering
    \includegraphics[width=0.5\textwidth]{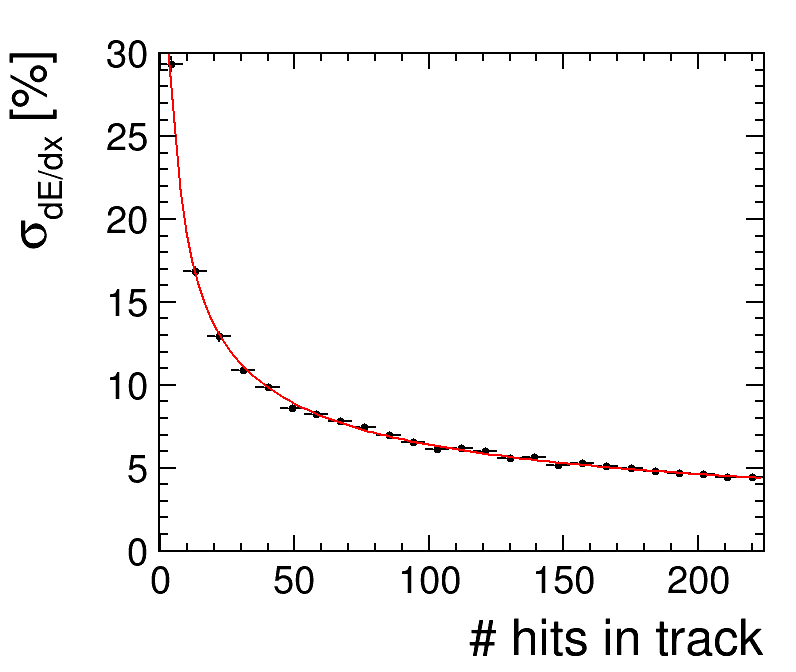}
    \caption{{\dEdx} resolution obtained using pseudo-tracks of various length.
        The red line represents the power\-/law fit described in the text.
    }
    \label{fig:dedxresolution}
 \end{figure}

The final goal of this analysis is to estimate the \dEdx resolution of the full size ILD TPC.
Therefore, the results need to be extrapolated to the greater track length, i.e.\ the greater number of possible hits.
Naively, one could expect the resolution to scale with the number of hits $N$ as $1/\sqrt{N}$ but experience of previous experiments shows, that it rather follows a power\-/law
\begin{equation*}
    \sigma_{\dEdx} \propto N^{-k}
    \label{eqn:dEdxScaling}
\end{equation*}
with an exponent $k$ in the range of \numrange{0.43}{0.48}, depending on experiment and \dEdx calculation methods~\cite{Blum:2008}.
This is caused by the Laundau\-/like shape of the probability distribution of the primary ionization process.
Therefore, the exponent needs to be determined for this setup individually.
Since the maximum number of valid hits in the prototype TPC is limited, a simple fit of the power\-/law to the resolution for tracks up to that length would introduce large uncertainties in the extrapolation.
Instead, randomly selected hits from multiple tracks are combined to form pseudo\-/tracks of arbitrary length.
Since the software processes events sequentially, this process is limited to hits from consecutive events~\cite{MarlinTPC,lcsoft,Marlin}.
To avoid biases from this method, care is taken that each hit is used only in a single pseudo\-/track.
The mean \dEdx value for each pseudo\-/track is then calculated as above.
Samples of pseudo\-/tracks with various numbers of hits are created and the resolution calculated for each.
\Cref{fig:dedxresolution} shows the resulting resolution versus the chosen track length.
The power\-/law fit returns an exponent of $k = \num{0.47(1)}$.
The \dEdx resolution at 220 hits as calculated from the fit is equal to $ \sigma_{\dEdx}= \SI{4.4(1)}{\percent}$, fulfilling the goal given in~\cite{ilc_tdr_detectors}.

\section{Impact of High Granularity on {\dEdx} resolution}
Simulation studies have been performed in order to investigate and quantify the impact of high granularity on the {\dEdx} resolution.

\subsection{The Cluster Counting Method}
\begin{figure}[tb]
    \centering
    \includegraphics[width=0.6\textwidth,keepaspectratio=true]{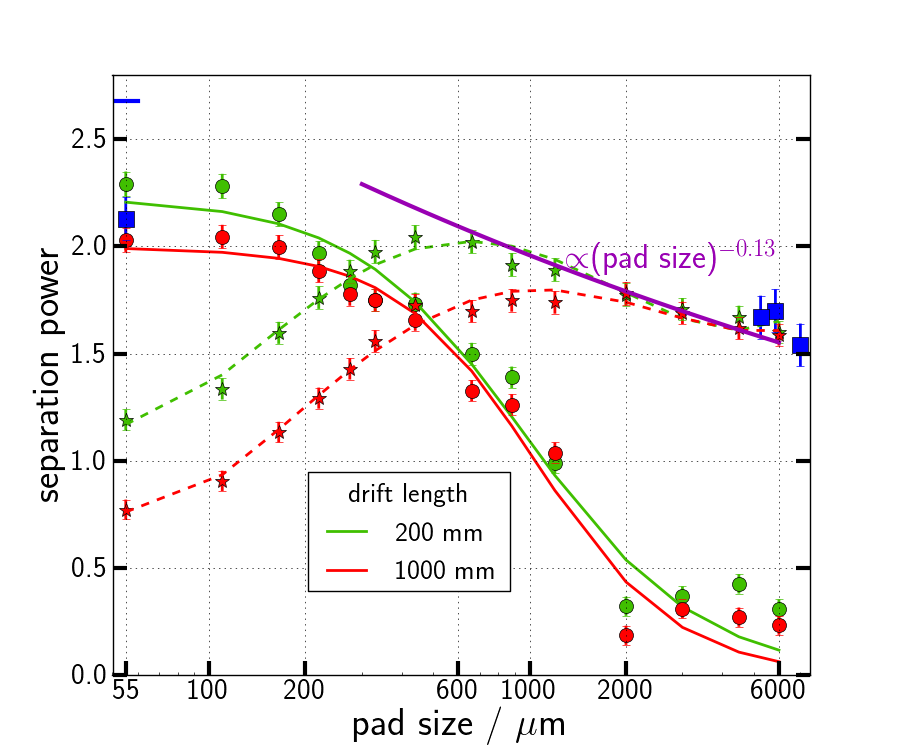}
    \caption{Pion\-/kaon separation power as a function of the pad size. Blue squares represent results from test beam measurements; stars and circles represent the simulations using the charge summation and cluster counting methods, respectively.}
    \label{fig:separationpowerVSpadsize}
\end{figure}

Traditionally, the energy loss is computed by summing all the electrons generated from ionization by the incident particle. The number of electrons produced in the ionization process is distributed according to a Landau distribution, which has a long tail towards large values. The relatively large width of the distribution is reflected in a degradation of the correlation between the measured energy and the momentum of the particle. An alternative method consists in counting the number of ionizing interactions of the incident particles. This follows a Poisson distribution with a significantly smaller width, compared to the Landau, therefore providing a better correlation and particle identification power~\cite{DEDXCLUSTER}. A comparison between the two methods has been performed studying the pion\-/kaon separation power as a function of the pad size, as shown in \cref{fig:separationpowerVSpadsize}, where the stars and the circles represent the simulation performed using the charge summation method and the cluster counting method, respectively. They are compared to results from test beam measurements represented by the blue squares. Below a pad size of about \SI{500}{\micro\m} the cluster counting method performs better than the charge summation method.

\subsection{The ROPPERI System}
In order to exploit the cluster counting capability, a high granularity readout system is needed. The ROPPERI (Readout Of a Pad Plane with ElectRonics designed for pIxels) system represents a possibility to combine high granularity, flexibility and large anode coverage of the pad-plane readout~\cite{ROPPERI2017}.

\begin{figure}[bp!]
    \centering
    \includegraphics[width=0.6\textwidth,keepaspectratio=true]{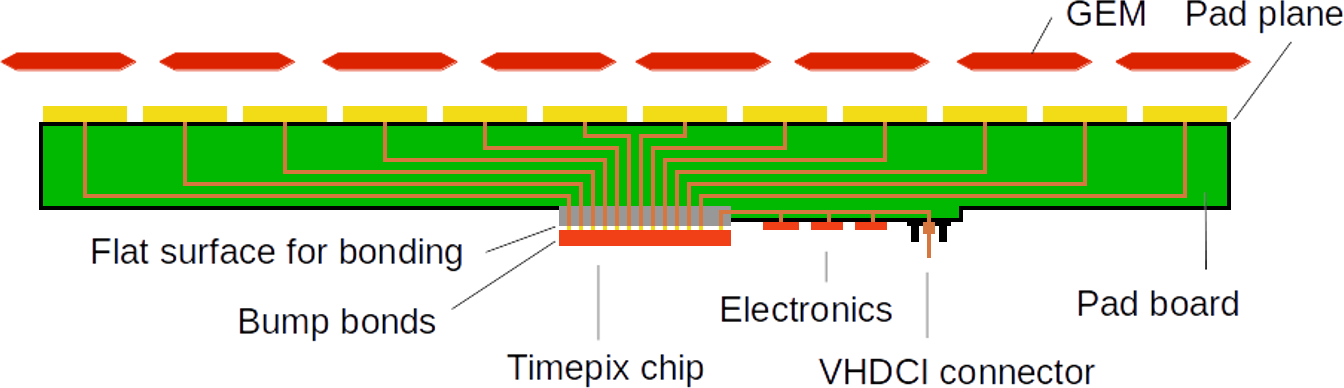}
    \caption{The ROPPERI board.
      \label{fig:ROPPERI}}
\end{figure}

The idea of the ROPPERI board implementation is sketched in \cref{fig:ROPPERI}. GEMs are used for amplification, small pads on a PCB form the anode and they are read out by a Timepix ASIC, which has a matrix of $256 \times 256 = 65536$ pixels with a \SI{55}{\micro\m} pitch. The connections from the pads are routed through the PCB to the ASIC which is bump bonded to the PCB surface. The Timepix power and communication pads are on the same side of the ASIC as the pixels, which makes it necessary to also connect them via bump bonds back into the PCB, where they are routed to a macroscopic connector plug (VHDCI).

A prototype board had been designed to study different pad sizes and line lengths, and the influence of the capacitance on the signal to noise ratio. Several boards have been produced out of which seven have been bump bonded in different tries. Out of these seven boards, only three worked completely. During the data taking, different thresholds have been used in runs with \numrange{100}{200} frames. The bump bonding tests showed issues connected to temperature during and after the bump bonding. Due to temperature difference and differing expansion coefficients of the materials of the chip and PCB, the bump connections broke. Until this happened, only a limited number of test runs were possible.

In the test runs, the worst pads and lines were about 11 times noisier than the bare Timepix chip with a known equivalent noise charge of about 90 electrons. So, under the assumption of a noise level of about 1000 electrons and \numrange{1e4}{2e4} signal electrons arriving on a pad after the GEM amplification, a signal to noise ratio higher than 10 can be estimated.

\section{The Large Prototype TPC Field Cage}
\begin{figure}[!bp]
  \begin{subfigure}[t]{0.4\textwidth}
    \centering
    \includegraphics[width=0.9\textwidth,height=0.33\textheight,keepaspectratio=true,clip,trim=0 65mm 15mm 20mm]{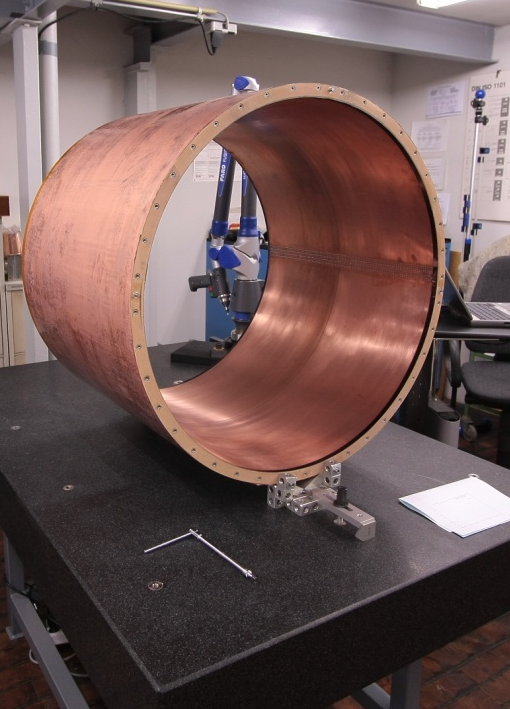}
    \caption{The current prototype field cage.}
    \label{fig:fielcageproto}
  \end{subfigure}%
  \begin{subfigure}[t]{0.6\textwidth}
    \centering
    \includegraphics[width=\textwidth,height=0.33\textheight,keepaspectratio=true]{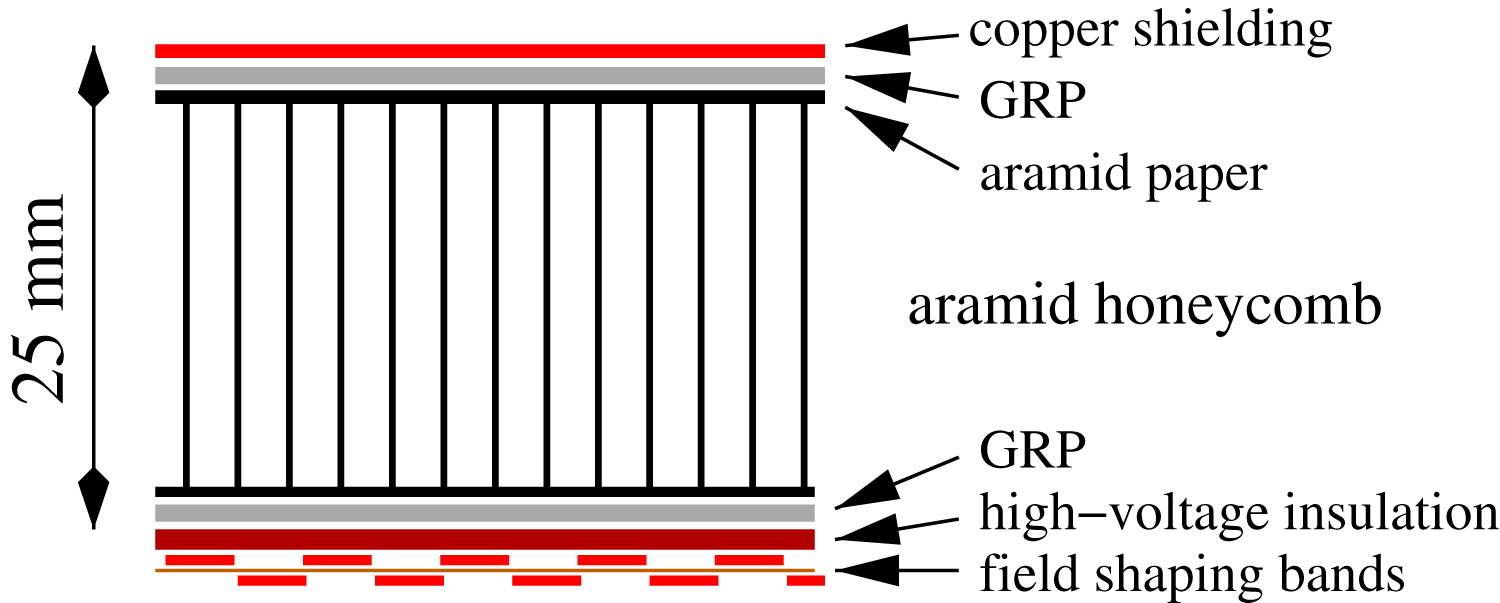}
    \caption{Sketch of the structure of the current field cage wall~\cite{SCHADE2011128}.}
    \label{fig:honeycomb}
  \end{subfigure}%
  \caption{The existing large prototype TPC field cage.}
\end{figure}

DESY is working on the construction of a new large prototype TPC field cage. A high precision TPC field cage for the ILD TPC has to guarantee a low material budget, a high voltage stability and high mechanical precision. Since the current large prototype field cage, shown in \cref{fig:fielcageproto}, does not fully meet the requirements on the mechanical precision~\cite{Schade2010,SchadePhD}, it was decided to build a new one. This field cage is build in the DESY workshop, so it also represents an opportunity to gain in\-/house experience in the design and construction of such a composite structure. The prototype is supposed to fit into the PCMAG, therefore its dimensions are restricted by the magnet geometry, as was the case for the previous field cage. The outer diameter is foreseen to be \SI{77}{\cm} leaving a gap of \SI{4}{\cm} to the inner wall of the magnet, which has a diameter of \SI{85}{\cm}.

As stated above, the \SI{61}{\cm} long field cage should be lightweight and at the same time mechanically very stable. Therefore, its structure is made of composite materials, already used in the construction of previous prototype field cages. \Cref{fig:honeycomb} shows a sketch of the wall structure, which is estimated to contribute \SI{1.31}{\percent} of a radiation length $X_0$.
An optimized design of field strips and mirror strips, implemented on a double\-/sided copper\-/coated polyimide foil, is used to generate a homogeneous field inside the cage.
A resistor chain linearly degrades the voltage along the length of the field cage, creating a constant field.
At the time of the construction of the current field cage, no producer was found to manufacture the \SI{61 x 226}{\cm} large, two\-/sided field strip foil in one piece. Therefore, it was produced from two about \SI{30}{\cm} wide foils which were connected during the construction. Now, the production of such a large foil in one piece is possible and a set of foils is being produced at the CERN workshop.

\begin{figure}[tbp]
  \begin{subfigure}[t]{0.5\textwidth}
    \centering
    \includegraphics[width=\textwidth,keepaspectratio=true]{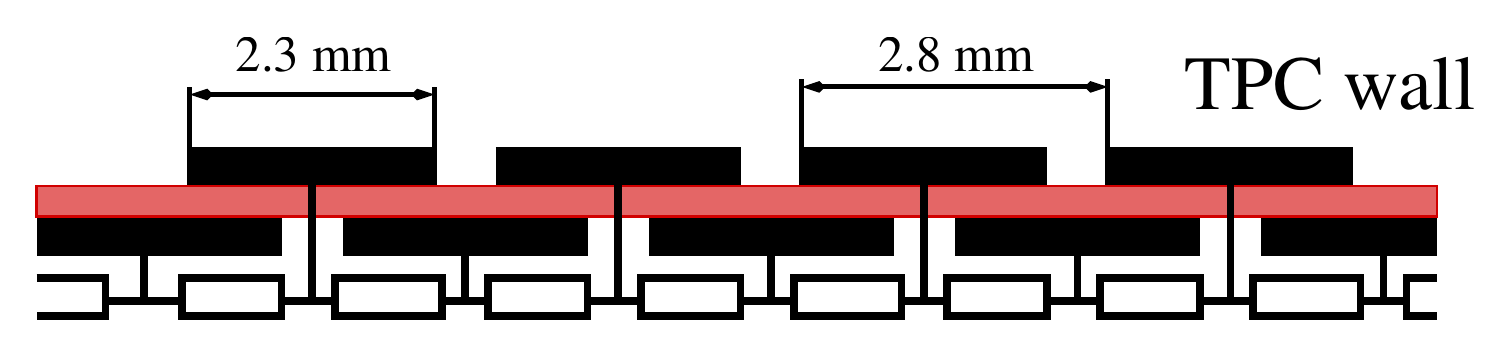}
    \caption{}
      \label{fig:afieldstripfoil}
  \end{subfigure}%
  \begin{subfigure}[t]{0.5\textwidth}
    \centering
    \includegraphics[width=\textwidth,keepaspectratio=true,trim=0 4.8mm 0 0]{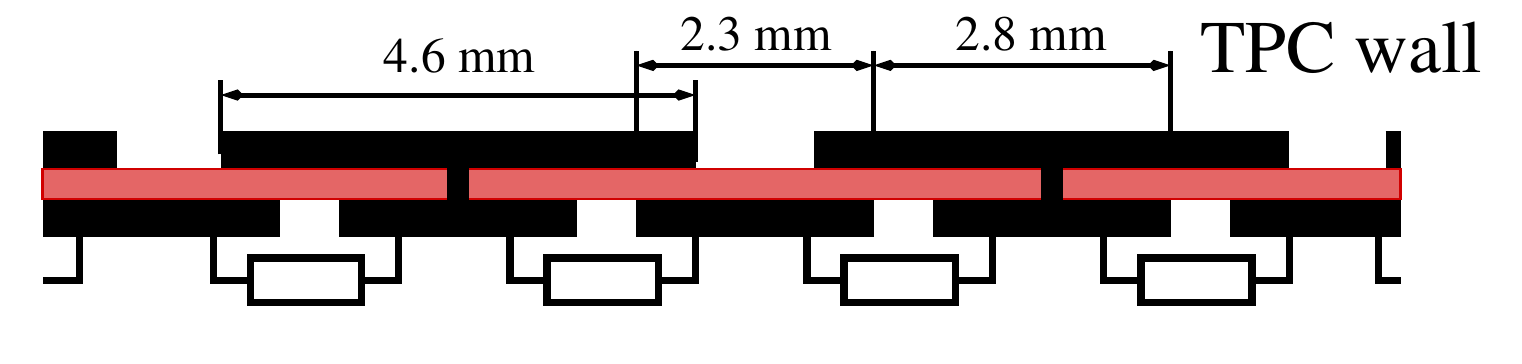}
    \caption{}
      \label{fig:bfieldstripfoil}
  \end{subfigure}
  \caption{Schematic view of cuts through the field strip foils perpendicular to the field strips.
    \subref*{fig:afieldstripfoil}~shows the current and
    \subref*{fig:bfieldstripfoil}~the new design.
    The field strips are depicted in black, the polyimide foil in red, the white boxes represent the resistors. Figures from~\cite{SchadePhD}.
  }
  \label{fig:fieldstripfoils}
\end{figure}

After the construction of the current field cage, the design studies of the field strip layout have been carried on and a simpler design delivering a comparable field quality was found. In the new design, each mirror strip spans over two gaps so only half of the resistors and vias are needed compared to the previous design. Details can be found in~\cite{SchadePhD}. \Cref{fig:fieldstripfoils} shows sketches of the cross sections of the old and new designs of the field strips.

\section{The LYCORIS Beam Telescope} \label{section:beamtelescope}
\begin{figure}[!bp]
    \centering
    \includegraphics[width=0.6\textwidth,keepaspectratio=true]{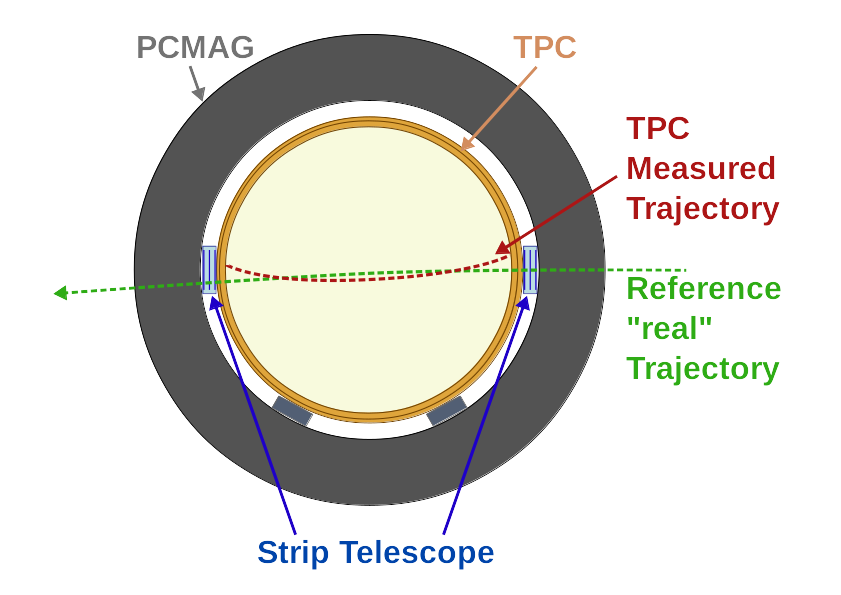}
    \caption{Sketch of the functionality of the LYCORIS telescope (here denoted by ``strip telescope'') installed within the PCMAG solenoid together with a large TPC prototype under test.
      \label{fig:sketch_telescope}}
\end{figure}

The LYCORIS Large Area Silicon Strip Telescope~\cite{LYCORISIEEE8824667,LYCORISVCIWU2019162864} is foreseen to be installed within the PCMAG at the DESY II Test Beam Facility.
The system requirements described here are for the use case of a TPC within the PCMAG. The telescope is designed to provide a precise reference measurement of the particle trajectory, as shown in the sketch in \cref{fig:sketch_telescope}. These reference measurements can be used to study and correct for potential inhomogeneities of the electric field within the TPC volume, which otherwise limit the achievable momentum resolution.

To meet these criteria, the telescope needs to provide a spatial single point resolution of better than $\sigma_\mathrm{bend} = \SI{10}{\micro\m}$ along the bending direction within the PCMAG and a resolution better than $\sigma_\mathrm{drift} = \SI{1}{\mm}$ along the drift axis of the TPC.

Hybrid-Less silicon strip sensors, designed by SLAC, are used for the LYCORIS telescope. They are characterized by an active area of \SI{10 x 10}{\cm} with a strip pitch of \SI{25}{\micro\m} and contribute \SI{0.3}{\percent} of a radiation length. The sensors are read out by an integrated digital readout chip called KPiX, also designed at SLAC, which is directly bump bonded onto the sensors and routed to the strips within the silicon sensor.

\begin{figure}[tbp]
    \centering
    \includegraphics[width=0.7\textwidth,keepaspectratio=true]{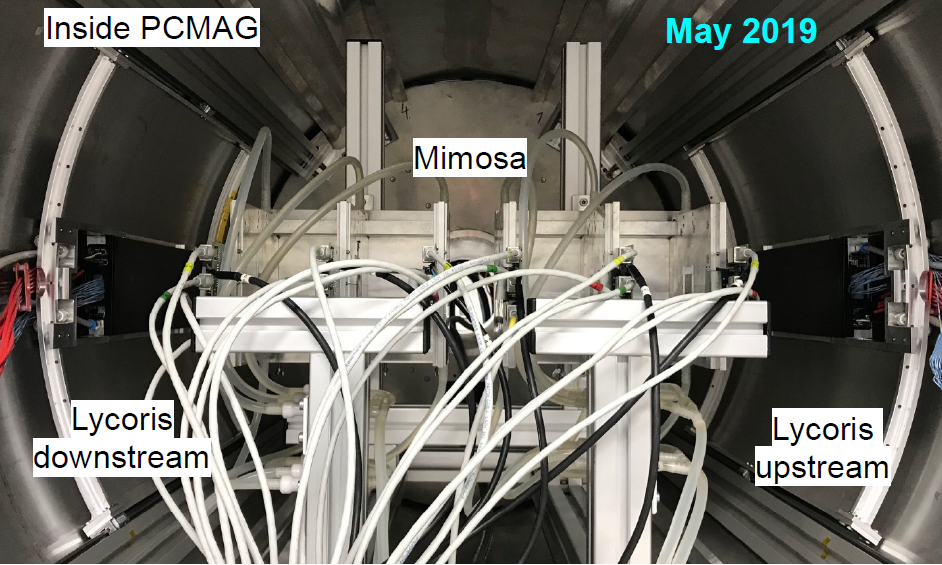}
    \caption{The LYCORIS telescope mounted in the PCMAG solenoid together with a Mimosa telescope.
      \label{fig:LycorisPlusMimosa}}
\end{figure}

The telescope has been fully assembled and successfully tested in several test beam campaigns, also in combination with EUDET-type Mimosa pixel telescopes~\cite{Jansen:2016bkd}. A picture of the LYCORIS telescope, used in combination with a Mimosa telescope inside the PCMAG solenoid is shown in \cref{fig:LycorisPlusMimosa}.
The system is undergoing final adjustments to be integrated into the infrastructure of the DESY II Test Beam Facility.

\section{Summary}
The DESY GridGEM modules have been shown to be able to fulfill the requirements set in~\cite{ilc_tdr_detectors} for several important parameters: point resolution, double\-/hit resolution and \dEdx resolution.
Simulation studies have shown that the \dEdx resolution may be improved with an optimized pad size of \orderof{\SI{500}{\um}}.
The ROPPERI system indicates a possible way to achieve such readouts with very high granularity.

A new, improved field cage for the large TPC prototype is in preparation.
The Lycoris beam telescope has been fully assembled and was successfully tested.
Some more work is needed to enable integration into the infrastructure of the DESY II Test Beam Facility.

\section{Acknowledgments}
\Cref{fig:LPTPC} is used with permission from the LCTPC Collaboration as authors of the article ``A time projection chamber with GEM-based readout'', Nuclear Instruments and Methods in Physics Research A 856 (2017) 109–118, Elsevier, \textcopyright{} 2016.\\
This material is based upon work supported by the National Science Foundation under Grant No.~0935316 and was supported by JSPS KAKENHI Grant No.~23000002. The research leading to these results has received funding from the European Commission under the 6th Framework Programme ``Structuring the European Research Area'', contract number RII3-026126, and under the FP7 Research Infrastructures project AIDA, grant agreement no.~262025.\\
Special thanks go to Y.~Makida, M.~Kawai, K.~Kasami and O.~Araoka of the KEK IPNS cryogenic group, and A.~Yamamoto of the KEK cryogenic centre for their support in the configuration and installation of the superconducting PCMAG solenoid.\\
The measurements leading to these results have been performed at the Test Beam Facility at DESY Hamburg (Germany), a member of the Helmholtz Association.\\
The authors would like to thank the technical team at the DESY II accelerator and test beam facility for the smooth operation of the test beam and the support during the test beam campaign. The contributions to the experiment by the University of Lund, KEK, Nikhef and CEA are gratefully acknowledged.

\printbibliography
\end{document}